\documentclass{kluwer}

\usepackage{epsfig}

\begin{document}
\begin{article}
\begin{opening}
\title{HST/STIS observations of sdBV stars: testing diffusion and pulsation theory}            

\author{S.~J. \surname{O'Toole}\email{otoole@sternwarte.uni-erlangen.de}}
\author{U. \surname{Heber}}
\institute{Dr Remeis-Sternwarte, Astronomisches Institut der Universit\"at
Erlangen-N\"urnberg, Sternwartstr.\ 7, Bamberg D-96049, Germany}                               
\author{P. \surname{Chayer}}
\institute{Dept. Physics and Astronomy, Johns Hopkins University, 3400
N. Charles St., Baltimore MD 21218, USA}

\author{G. \surname{Fontaine}}
\institute{D\'epartement de Physique, Universit\'e de Montr\'eal, CP 6128,
Station Centreville, Montr\'eal, QC H3C-3J7, Canada}

\author{D. \surname{O'Donoghue}}
\institute{South African Astronomical Observatory, PO Box 9, Observatory 7935,
South Africa}

\author{S. \surname{Charpinet}}
\institute{Observatoire Midi-Pyr\'en\'ees, 14 Avenue E. Belin, 31400 Toulouse, France}

\runningtitle{HST/STIS Observations of sdBV stars}
\runningauthor{S.~J. O'Toole et al.}

\begin{abstract} 
We present the initial results of an abundance analysis of echelle UV
spectra of 5 hot subdwarf B (sdB) stars. These stars have been
identified as core helium burning objects on the extreme Horizontal
Branch. Around 5\% of sdBs show short-period acoustic mode oscillations.
Models predict that these oscillations are due to an opacity bump
caused by the ionisation of iron group elements. The necessary metal
abundance has to be maintained by diffusive equilibrium between
gravitational settling and radiative levitation. However, analyses of
high resolution optical spectra has revealed that we cannot discriminate
between pulsating and non-pulsating sdBs on the basis of the surface
iron abundance. We have therefore obtained HST/STIS observations of 3
pulsators and 2 non-pulsators in the near and far UV to measure the
surface abundance of elements that are unobservable from the ground. The
overall aim of our study is to test diffusion and pulsation
calculations by searching for significant differences between these
surface abundances.
\end{abstract}


\end{opening}

\section{The problem}

Models of pulsating subdwarf B stars suggest that the pulsations are
driven by an opacity bump due to iron and other metallic species at
temperatures of $\sim2\times10^5$\,K in the sdB envelope (see
Charpinet et al. 2001 for a review). The models require a sufficient
abundance of Fe at an appropriate level in the envelope, an abundance
which \emph{can} be maintained by diffusive equilibrium between
graviational settling and radiative levitation.
While these models give rise to pulsations in
hot sdBs, they do not yet explain why, when given two
spectroscopically similar stars, one will pulsate and the other will
not. This overlap in the $\log g-T_{\mathrm{eff}}$ plane was
demonstrated by Koen et al.\ (1999). Fontaine \& Chayer (1998) suggest
different mass-loss rates as one possible explanation, whereby a star
with a higher mass-loss rate may have a depleted Fe abundance, and
therefore not show pulsations.

The goal of this project is to investigate the metal abundances of 3
pulsating sdBs and 2 non-pulsating sdBs using HST/STIS UV
spectra, and to determine whether or not the latter stars' abundances
are significantly different from those of the former stars. With these
observations we can derive abundances for elements not observable from
the ground, e.g.\ the iron group. If there are no significant
differences between the abundances patterns, then this may suggest
that the stellar winds suggested by Fontaine \& Chayer are the
discriminating factor. The second (but no less important) goal of the
project is to compare detailed abundances in all five stars with the
predictions of diffusion theory.

\section{The targets}
Observations were made using the Space Telescope Imaging Spectrograph
(STIS) onboard HST. We used medium
resolution gratings in echelle mode in both the far UV (grating E140M)
and the near UV (grating E230M). The FUV spectra contain 42 useful
orders with wavelength range 1149-1730\,\AA, while the NUV contain 40
orders with wavelength range 1635-2365\,\AA. Of the 3 pulsators
observed, PG\,1219+534 and Feige 48
were observed for two orbits in both the far UV and near UV, while
PG\,1605+072 was observed twice in the far UV, but only once
in the near UV. Ton S-227 and Feige 66 were observed once in the far
and once in the near UV. These two stars do not show pulsations, but
are spectroscopic twins of PG\,1219+534.

\begin{figure}[htbp]
\begin{center}
\epsfig{file=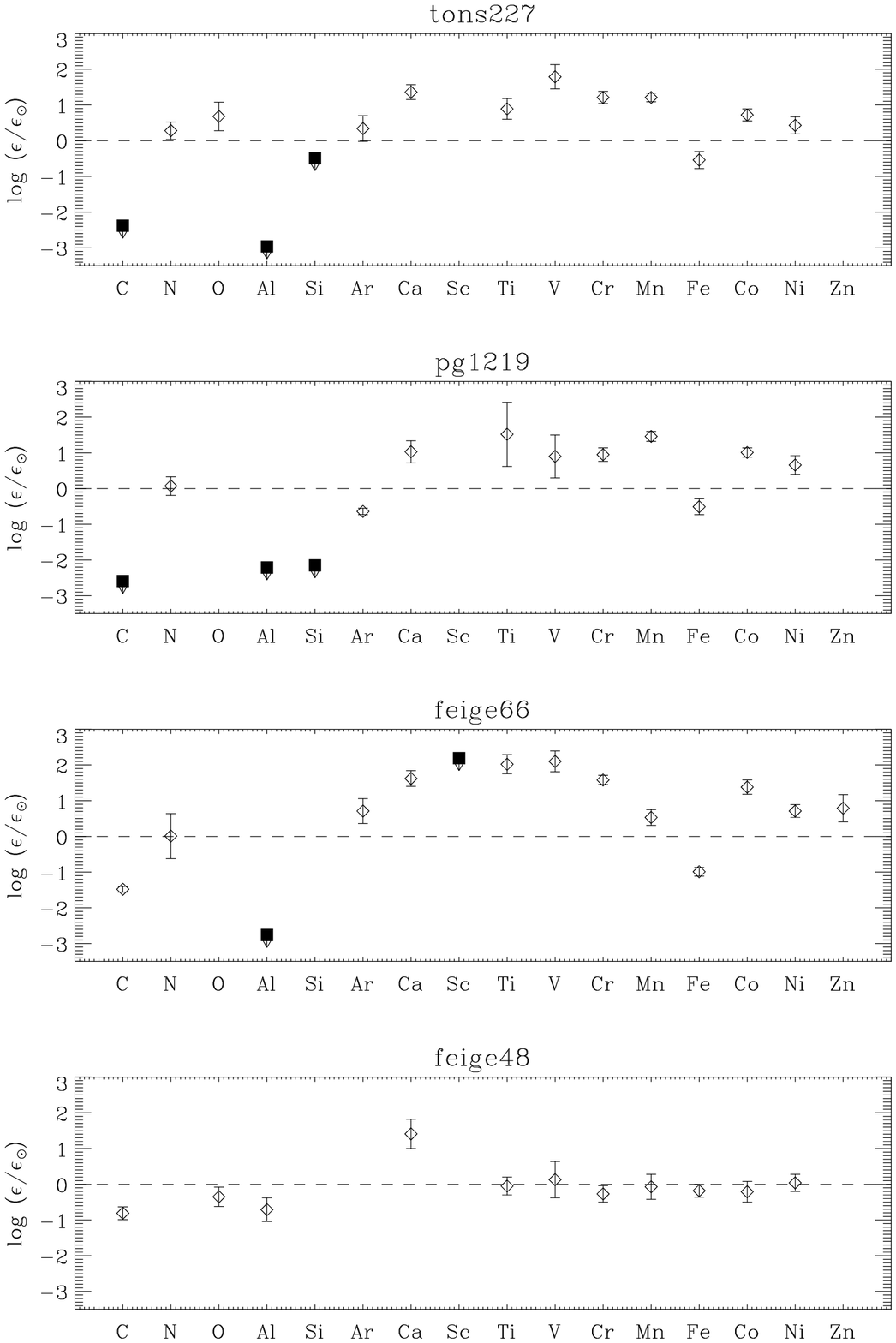,scale=0.44}
\end{center}
\caption{Abundances for four sdB stars, derived from near UV
HST/STIS spectra.}
\label{fig:NUVabund}
\end{figure}

\section{Initial results: metal abundances in the NUV}

Due to the large number of metal lines, it was difficult to determine
the continuum level for each spectrum, meaning that the errors for
each abundance presented may be underestimated. In order to overcome
the substantial line blending in the spectra, we used a solar
metallicity model and adjusted the abundances of each element until
$\chi^2$ was minimised. This
was done using Michael Lemke's version of the LINFOR program
(originally developed by Holweger, Steffen, and Steenbock at Kiel
University). Oscillator strengths were taken from the Kurucz line
list, as were damping constants for all metal lines. For all of the
results presented here, the microturbulent velocity was set to
0\,km\,s$^{-1}$. The current study is limited to the near UV spectra,
since for these spectra it was easier to determine a continuum
level. Abundances will be derived from the far UV spectra and for
PG\,1605+072 (whose spectra are complicated by rapid rotation) in the
near future. The results for the four studied stars are shown in
Figure \ref{fig:NUVabund}.

From our preliminary analysis, we find no significant differences
between abundances of pulsating and non-pulsating sdBs. In particular
the abundances of PG\,1219+534 and Ton S-227 are remarkably
similar. There is a noticeable difference between the abundance
pattern of Feige 48 and the other 3 stars. The hotter, higher surface
gravity stars all show supersolar iron group abundances, while for
Feige 48 these elements have solar abundances. The stand out element
is calcium, which is supersolar in all 4 stars. These effects appear
to be due to ``selective diffusion'', i.e.\ the interplay between
radiative levitation and gravitational settling. It is yet to be seen
whether diffusion models can explain these effects.

\section{Comparisons, conclusions and future work}

We can compare the results presented here with previous work by Heber
et al.\ (2000) and Baschek et al.\ (1982). For Feige 48 and
PG\,1219+534, Heber et al.\ measured (or set upper limits) for
6 species we have measured from high-resolution spectra taken with
the Keck telescope. A slightly lower Fe abundance is the only
noticeable difference between our results and that of Heber et al. All
other elements are in reasonable agreement. This lower Fe abundance is
also seen in FUSE spectra (Chayer, these proceedings). 
Because of the relative inaccuracy of the atomic data available to the
Baschek et al.\ (1982) analysis of Feige 66 using high-resolution IUE
spectra, it is difficult to compare their results with ours. However,
we can say that, apart from calcium, the agreement appears to be
reasonably good. Overall we believe that our analysis is consistent
with the analyses of Heber et al.\ (2000) and Baschek et al.\ (1982).

All of the high $\log g$ stars have supersolar iron group
element abundances, and in particular PG\,1219+534 and Ton S-227
appear to be very similar, while Feige 48, with a substantially lower
$\log g$, has approximately solar abundances. Do these results present
a problem for the Charpinet et al. Fe driving model? At this stage we
do not want to draw any premature conclusions: more abundance analyses
of pulsating and non-pulsating sdBs are needed to see any heavy
element abundance patterns (e.g. Chayer et al., in preparation). High
resolution optical and FUV observations or more sdB stars should
clarify the driving mechanism.

\section{References}

Baschek B., Hoflich P., \& Scholz M., 1982, \emph{A\&A} 112, 76

\noindent Charpinet S., Fontaine G., \& Brassard P., 2001, \emph{PASP} 113, 775

\noindent Fontaine G. \& Chayer P., 1997, in: \emph{3rd Conf. on Faint
  blue stars}, Schenectady, L. Davis press, p.169

\noindent Heber U., Reid I.~N. \& Werner K., 2000, \emph{A\&A} 363, 198

\noindent Koen C., O'Donoghue D., Kilkenny D., Stobie R.~S. \& Saffer R.~A.,
1999, \emph{MNRAS} 306, 213

\end{article}
\end{document}